\documentclass[12pt,a4paper]{article}
\usepackage{graphicx}

\newcommand{\tw}{0.95\textwidth}
\newcommand{\htw}{0.47\textwidth}
\newcommand{\eq}[1]{(\ref{eq:#1})}

\begin{document}

{\bf \begin{center}
{\Large \bf
The results of ATIC-2 experiment for elemental spectra of cosmic rays
 }\\
\vspace{5mm}
{
\bf
A.D. Panov\footnote[1]{Skobeltsyn Inst. of Nuclear Physics, Moscow State University, Moscow, Russia}, 
J.H. Adams, Jr.\footnote[2]{Marshall Space Flight Center, Huntsville, AL, USA},
H.S. Ahn\footnote[3]{University of Maryland, Institute for Physical Science \& Technology, College Park, MD, USA},
G.L. Bashindzhagyan\footnotemark[1], 
K.E. Batkov\footnotemark[1], 
J. Chang\footnote[4]{Purple Mountain Observatory, Chinese Academy of Sciences, China}${}^,$\footnote[5]{Max-Planck Institut for Solar System Research, Katlenburg-Lindau, Germany},
M. Christl\footnotemark[2],
A.R. Fazely\footnote[6]{Southern University, Department of Physics, Baton Rouge, LA, USA},
O. Ganel\footnotemark[3],
R.M. Gunashingha\footnotemark[6],
T.G. Guzik\footnote[7]{Louisiana State University, Department of Physics and Astronomy, Baton Rouge, LA, USA},
J. Isbert\footnotemark[7],
K.C. Kim\footnotemark[3],
E.N. Kouznetsov\footnotemark[1],
M.I. Panasyuk\footnotemark[1], 
W.K.H. Schmidt\footnotemark[5],
E.S. Seo\footnotemark[3],
N.V. Sokolskaya\footnotemark[1],
John W. Watts\footnotemark[2],
J.P. Wefel\footnotemark[7],
J. Wu\footnotemark[3], and
V.I. Zatsepin\footnotemark[1], 
}
\end{center}
}

\begin{abstract}
The results of ATIC-2 baloon experiment (2002--2003) for energy
spectra of protons, He, C, O, Ne, Mg, Si, Fe, some groups of nuclei,
and all-particle spectrum in primary cosmic rays are presented in
energy region 50~GeV--200~TeV. The conclusion is that the spectra of
protons and helium nuclei are essentially different (the spectrum of
protons is steeper)  and the spectra of protons and heavy nuclei have
no-power form.
\end{abstract}

\section{Introduction}

ATIC (Advanced Thin Ionization Calorimeter) is a balloon borne
experiment designed to measure energy spectra of cosmic rays with
individual charge resolution from protons to iron in the energy
interval approximately from 50~GeV to 200~TeV.  ATIC is comprised of a
silicon matrix, a carbon target, scintillator hodoscopes, and a fully
active BGO calorimeter. The detailed description of the spectrometer
is given in \cite{ATIC_GUZIC2004}, the method of measurement of
primary particle charge and the charge resolution achieved in the
spectrometer is described in \cite{ATIC_MSU2004A}.  A matrix of
silicon detectors that permits to solve the problem of albedo current
by the method of high segmentation of charge detector was used in the
high energy cosmic ray physics for the first time
\cite{ATIC_MSU2005A}.  ATIC had two successful balloon flights in
Antarctica: from 28 Dec 2000 to 13 Jan 2001 (test flight ATIC-1) and
from 29 Dec 2002 to 18 Jan 2003 (science flight ATIC-2).  The
preliminary results of the processing of the ATIC-2 flight data were
published in papers \cite{ATIC_MSU2004B,ATIC_MSU2006A}.


The calorimeter of the ATIC spectrometer is thin, that is only a part
of the energy of primary particle is detected by the calorimeter.
Therefore there exist a problem to reconstruct the spectrum of primary
particles from the spectrum of energy deposit.  The spectrum of energy
deposit $f(E_d)$ for the particles of some definite sort is connected
with the primary spectrum $\Phi(E_0)$ by the Fredholm equation of the
first kind:
\begin{equation}
  \label{eq:Fredholm}
  f(E_d) = \int A(E_d,E_0) \Phi(E_0) dE_0,
\end{equation}
where $A(E_d,E_0)$ is the apparatus function which produces for each
primary energy $E_0$ the distribution of the energies $E_d$ deposited
in the calorimeter.  To obtain the spectrum of primary energies one
must solve the equation \eq{Fredholm} for $\Phi(E_0)$ with known
function $f(E_d)$.  This problem belongs to the sort of ill-defined
reverse problems. Different approximate methods are used to solve such
problems.  The scaling approximation that is used in emulsion
calorimeters \cite{BURNETT86} was used in papers
\cite{ATIC_MSU2004B,ATIC_MSU2006A} for transition from the spectra of
energy deposit to the primary spectra.  It is supposed in such
approximation that the fraction of the energy deposited in the
calorimeter (the ratio $E_d/E_0$) and the effectiveness of
registration do not depend from the primary energy $E_0$. Therefore
the primary spectrum could be obtained from the energy deposit
spectrum simply by energy shift and normalization by the effectiveness
of registration.


A simulation of cascades propagation in the spectrometer show that
scaling approximation describes the operation of the device very
roughly.  The method of papers \cite{ATIC_MSU2004B,ATIC_MSU2006A}
should be improved.  Two more precise methods to solve the problem of
reconstruction the primary spectra \eq{Fredholm} are used in the
present work: direct solution of the Fredholm equation with Tikhonov's
regularization and the method of differential shifts which is a
generalization of the known method \cite{BURNETT86}.  Besides, a new
more precise calibration of the calorimeter and an improved correction
of the thermal sensitivity of calorimeter measurement tract were used
in present work.  Instead of using of mean flight temperature of the
device in temperature correction method as in papers
\cite{ATIC_MSU2004B,ATIC_MSU2006A}, instantaneous values of
temperature were accounted and used in new improved method of
correction of thermal sensitivity.


\section{Reconstruction of primary spectra}

The experimental differential spectra of energy deposited in the
calorimeter are recorded as a number of counts in
logarithmic-equidistant bins that divide the whole region of the
investigated energies onto $n$ parts.  If the primary spectra is
written in the same form then the integral equation \eq{Fredholm} in
discrete form may be written as a system of equation
\begin{equation}
  \label{eq:Discrete}
  M_i = \sum_{j = 1}^n a_{ij} N_j,\quad j = 1,2,\dots n,
\end{equation}
where the notations are obvious.  The search of the primary spectrum
$\{N_j, j=1,\dots,n\}$ based on the direct solution of discrete
Fredholm equation \eq{Discrete} with using of regularization leads to
minimization of some functional and using the Monte Carlo method for
calculation of statistical errors.  The realization of the method for
processing of the ATIC data was described in details in
\cite{ATIC_2005B}. 


It was shown that the method of solution of the reverse problem with
regularization for protons and helium nuclei leads to stable result. 
This means that if one processes a spectrum of energy by parts
then one obtains the respective parts of the primary spectrum that are
in the accordance with each other.  But it turned out that it was not
the case for nuclei $Z>=6$.  So the method of differential shifts that
generalizes the known method \cite{BURNETT86} based on exact scaling
was developed to process the spectra for nuclei $Z>=6$.  The energy
scaling factor is calculated separately for each value of $E_d$ in the
differential shifts method.  For the deposited energy value that is
exactly equal to the logarithmic mean of $i$-th bin of the spectrum
($E_i$) the primary energy is reconstructed by the formula
\begin{equation}
  \label{eq:DiffShift}
  E^{(i)}_0 = \frac{\sum_{j=i}^n a_{ij} E_j K(E_j)}%
                   {\sum_{j=i}^n a_{ij} K(E_j)},
\end{equation}
where $K(E_j)$ is an initial approximation for the primary spectrum. 
For intermediate values of $E_d$ the primary energy is estimated by
interpolation of the equation \eq{DiffShift}.  As an initial
approximation a power-form spectra with the integral index $\gamma =
1.6$ for momentum were used.


It was checked that the method of reconstruction of primary spectra by
solution of reverse problem with regularization and the method of
differential shifts produce the same (up to statistics) result in the
common for both methods energy region of the primary spectra.  But for
all, the method of differential shifts is stable for all species, though
the lower threshold for reconstructed spectra is approximately 200~GeV
instead 50~GeV in the method of regularization.  Another advantage of
the method of differential shifts upon direct solution of reverse
problem is that the first is usable in the region of data with low
statistics where the direct method is already unusable.  But it is needed
for the differential shifts method that the spectrum was not different
from the power one too much. It is the case for nuclei $Z \ge 6$.


The absolute fluxes of primary particles are determined by
normalization of the reconstructed primary spectra by geometrical
factor of the device, live time of the exposition, correction on the
residual atmosphere, and correction of effectiveness of algorithm of
trajectory reconstruction. 

\section{Results}

\begin{figure}[h]
  \begin{center}
    \includegraphics[width=\htw]{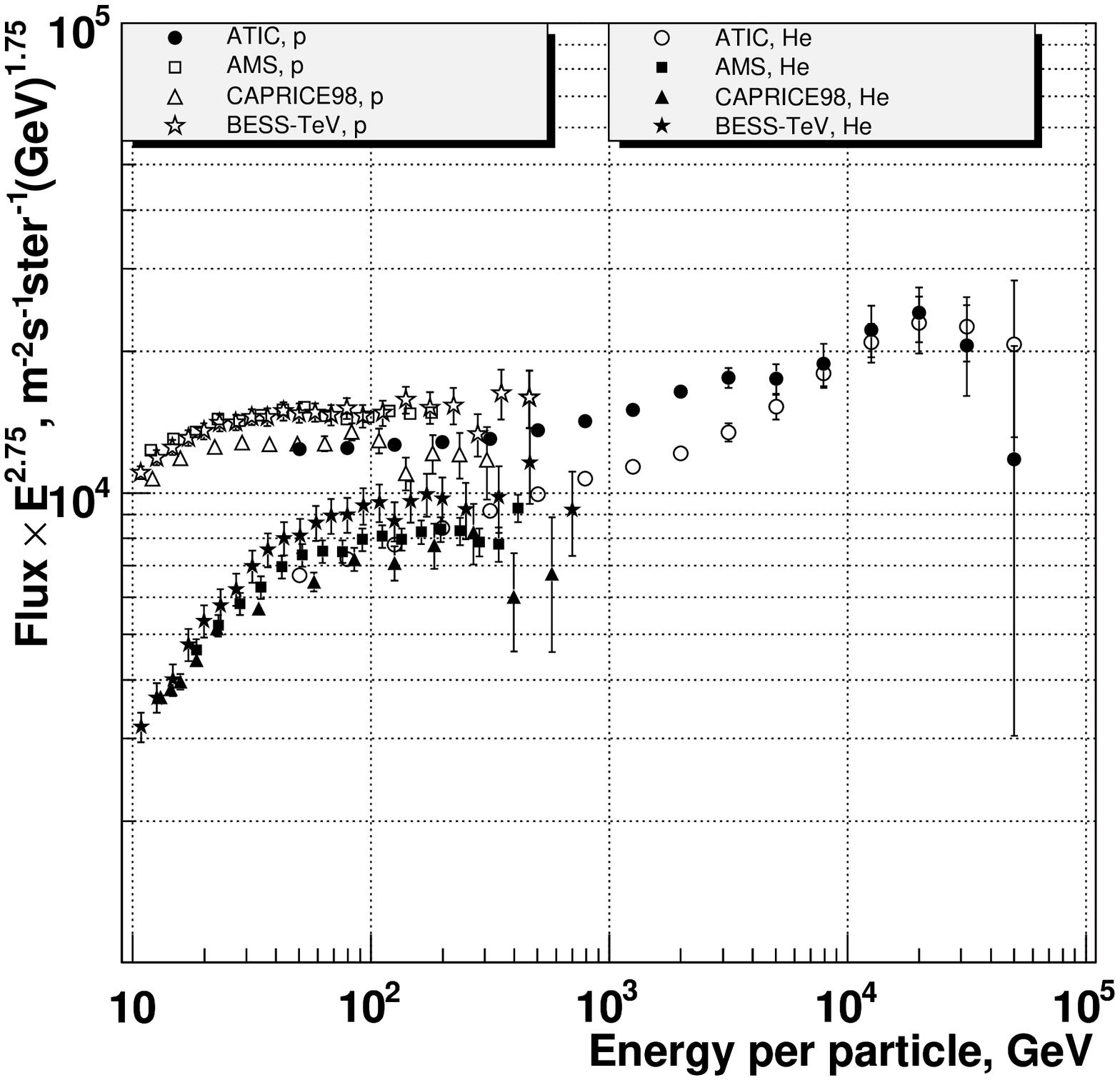}
    \includegraphics[width=\htw]{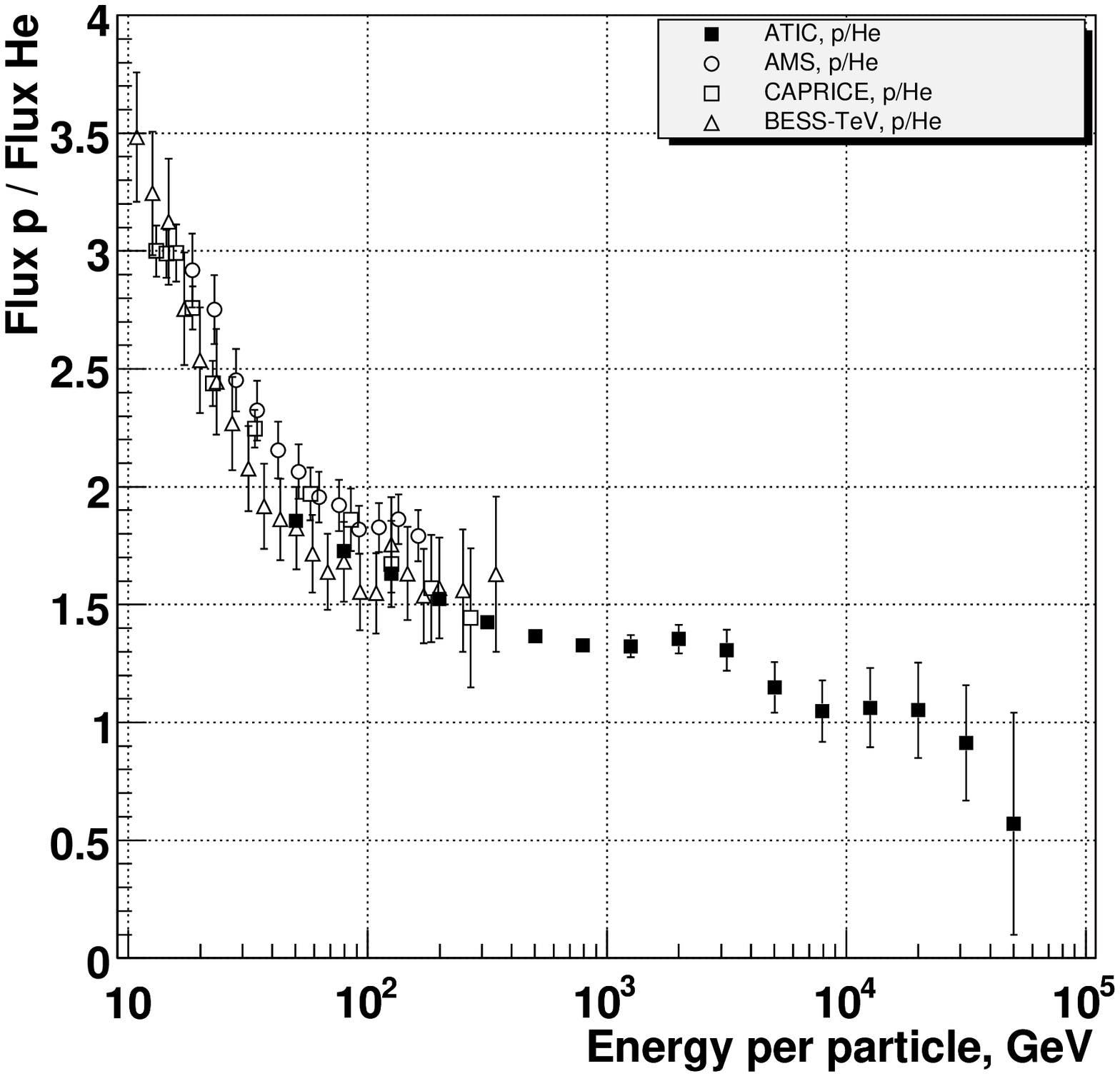}
  \end{center}
\caption{The spectra of protons and helium nuclei (at the left) and the ratio
of the pronon flux the helium flux (at the right)} 
\label{PHe}
\end{figure}

The obtained primary spectra of protons and helium nuclei (here and
everywhere in this paper in the terms of energy per particle) and the
ratio of the fluxes of protons and helium nuclei against the primary
energy are shown in fig.~\ref{PHe}.  The ATIC-2 results are compared
with the results of experiments AMS
\cite{AMS_ALCARAZ2000,AMS_ALCARAZ2000B}, CAPRICE-98
\cite{CAPRICE_BOEZIO2003}, BESS-TeV \cite{BESSTeV2004}.  The spectra
of protons and helium nuclei have definitely different slopes and are
in the accordance with the data of magnetic spectrometers
\cite{AMS_ALCARAZ2000}--\cite{BESSTeV2004} in the region of low
energies (where the data of magnetic spectrometers are defined).

\begin{figure}[h]
  \begin{center}
    \includegraphics[width=\tw]{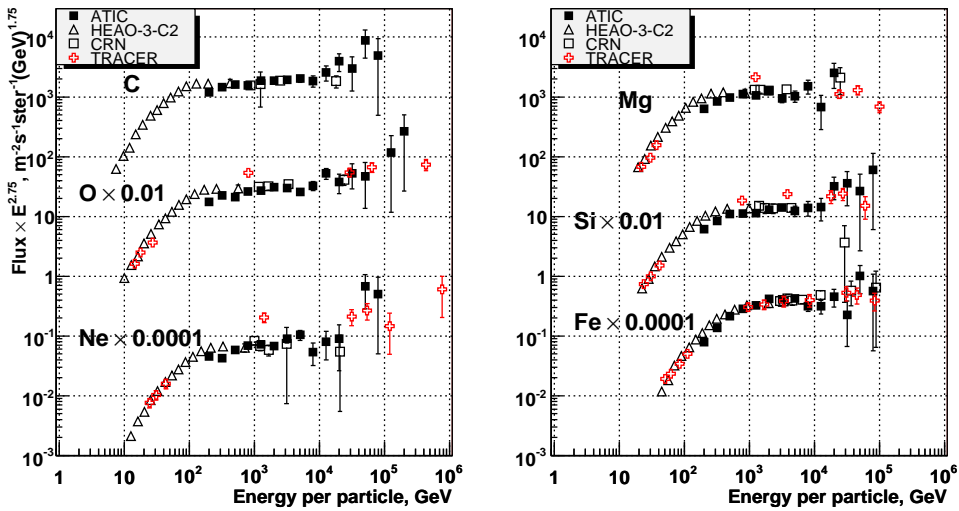}
  \end{center}
\caption{The spectra of even nuclei: C, O, Ne, Mg, Si, Fe.} 
\label{Nuclei}
\end{figure}

The primary spectra of some even nuclei $Z\geq 6$ are shown in
fig.~\ref{Nuclei}.  The data of ATIC-2 are compared with the results
of HEAO-3-C2 \cite{HEAO_ENGELMANN1990}, CRN \cite{CRN_MULLER1991},
TRACER (preliminary results) \cite{TRACER2005A}.  One can note a good
agreement except a little systematic exceeding of HEAO data over
ATIC-2 data in the energy region 200--700~GeV.

\begin{figure}[h]
  \begin{center}
    \includegraphics[width=\tw]{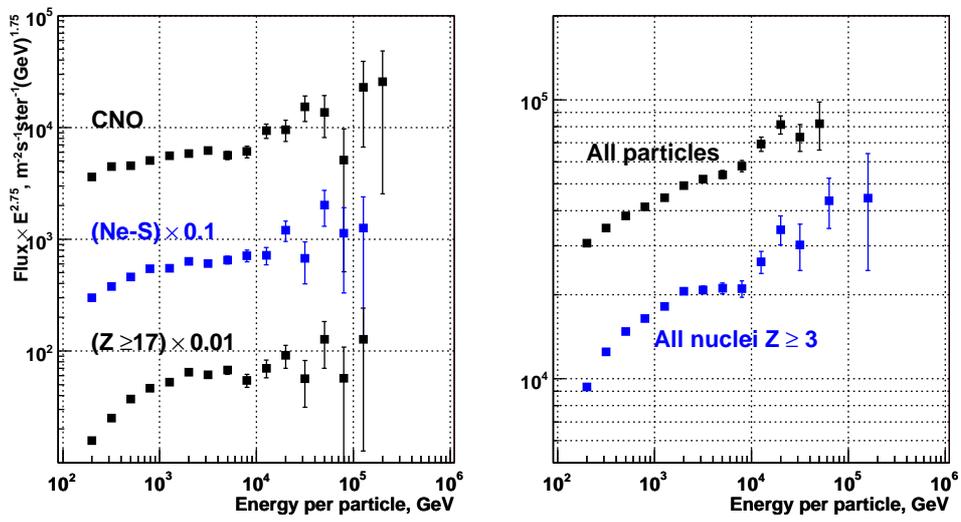}
  \end{center}
  \caption{The spectra of different nuclei groups and the spectrum of
    all particles. The upper threshold in the spectrum of all nuclei
    $Z\geq3$ is higher than in the spectrum of all particles because
    the spectra of protons and helium (see.  fig.~\protect\ref{PHe})
    can be reconstructed by the method of regularization only up to
    80~TeV.} 
\label{All}
\end{figure}

The spectra of different nuclei groups and the spectrum of all
particles from ATIC-2 data are shown in fig.~\ref{All}.  One can notes
a definite flating of spectra for nuclei heavier than helium in the
region of primary energies $E_0>10$~TeV. It especially clear in the
summary spectrum of nuclei $Z\geq3$ (fig.~\ref{All}) The shape of the
spectra of all particles is close to the power low, but in the light
of the results of ATIC-2 experiment this is seen as a random event
since the spectra of different nuclei groups (including the spectra of
protons and helium, fig.~\ref{PHe}) are rather far from a power law. 

The analysis of ATIC-2 data have been continuing. 


This work was supported in Russia by RFBR grats 02-02-16545,
05-02-16222; in USA by NASA grants NNG04WC12G, NNG04WC10G, NNG04WC06G;
J. Chang is grateful to Ministry of Science and Technology of China
for supporting by grant 2002CB713905. 


\end{document}